\documentclass[11pt]{article} 
\usepackage{graphicx}
\usepackage{amsmath,multirow}
\usepackage{enumerate,array}
\usepackage{amsfonts}
\usepackage{amssymb}
\usepackage{epsfig}
\usepackage{graphics}
\usepackage{multirow}
\usepackage{enumitem}
\usepackage{color}
\usepackage{lineno}
\usepackage{amscd}

\newcommand{\beq}{\begin{equation}}
\newcommand{\eeq}{\end{equation}}

\makeatletter
\renewcommand\@biblabel[1]{}
\makeatother

\renewcommand{\baselinestretch}{1.3}

\begin{document}
	
\title{Generalized logistic model for $r$ largest order statistics, with hydrological application} 

\author{Yire Shin$^{1}$$\;$ and $\;$Jeong-Soo Park$^{1,*}$  \\
	\small\it 1: Department of Mathematics and 
	Statistics, Chonnam National University, Gwangju 61186, Korea\\
	\small\it *: Corresponding author, email : jspark@jnu.ac.kr
}
\date{}
\maketitle 

\begin{abstract}
The effective use of available information in extreme value analysis is critical because extreme values are scarce. Thus, using the $r$ largest order statistics (rLOS)  instead of the block maxima  is encouraged. Based on the four-parameter kappa model for the rLOS (rK4D), we introduce a new distribution for the rLOS as a special case of the rK4D. That is the generalized logistic model for rLOS (rGLO). 
This distribution can be useful when the generalized extreme value model for rLOS is no longer efficient to capture the variability of extreme values. Moreover, the rGLO enriches a pool of candidate distributions to determine the best model to yield accurate and robust quantile estimates. We derive a joint probability density function, the marginal and conditional distribution functions of new model. The maximum likelihood estimation, delta method, profile likelihood, {order selection by the entropy difference test,} {cross-validated likelihood criteria,} and model averaging were considered for inferences. The usefulness and practical effectiveness of the rGLO are illustrated by the Monte Carlo simulation and an application to extreme streamflow data in Bevern Stream, UK. \textcolor{blue}{This paper was published at April, 2024: Shin, Y., Park, J-S. Generalized logistic model for $r$ largest order statistics, with hydrological application. Stoch Environ Res Risk Assess 38, 1567–1581 (2024). https://doi.org/10.1007/s00477-023-02642-7. In this revision, some modification and correction from the published one are made on sentences and formula with blue color.} 
\end{abstract}

\vspace{3mm} \noindent {\bf Keywords}:
Cross-validated likelihood; Entropy difference test; Flood frequency analysis; Generalized extreme value distribution; Log-logistic distribution; Return level.


\section{Introduction}
Extreme values are scarce by definition; thus, a model fitted to the block maxima may have a large variance, especially for estimating high quantiles of the underlying distribution. This issue has motivated two well-known generalizations. One uses the generalized Pareto model (Pan et al.~2022) based on exceedances of a high threshold, and the other is based on the $r$ largest order statistics (rLOS) within a block (Coles 2001). The latter is a compromise between the block maxima model and the generalized Pareto model (Rieder 2014). This study focuses on models for the rLOS statistics.

The Gumbel model for rLOS (rGD) was developed by Smith (1986), building on theoretical developments by Weissman (1978). It was extended to the generalized extreme value (GEV) model for rLOS by Tawn (1988), who recommended declustering the data before selecting the order statistics. We refer to the GEV model for rLOS as the rGEVD. Including more data up to the $r$th order statistics other than just one set of maxima  in each block may improve the precision of model estimation, and the interpretation of parameters is unchanged from the univariate ($r=1$ case) GEV distribution. The rGEVD has been employed in many applications (Dupuis 1997; Zhang et al.~2004; Soares and Scotto 2004; An and Pandey 2007; Wang and Zhang 2008; Aarnes et al~ 2012; Feng and Jiang 2015; Naseef and Kumar 2017).

The number $r$ comprises a bias-variance trade-off. Small values of $r$ consist of few data points leading to high variance, whereas large values of $r$ are likely to violate the asymptotic support for the model, leading to high bias (Coles 2001). The selection of $r$ is important in the model for rLOS. Smith (1986) and Zhang et al.~(2004) suggested $r=5$ as a reasonable compromise. Bader et al.~(2017) developed automated methods for selecting $r$ from the rGEVD. Serinaldi et al.~(2020) provided a theoretical rule to select $r$, which supported a popular rule of using $r=3$ in hydrology.

\begin{figure}[htb]
	\centering
	\begin{tabular}{l}	\includegraphics[scale = 0.4]{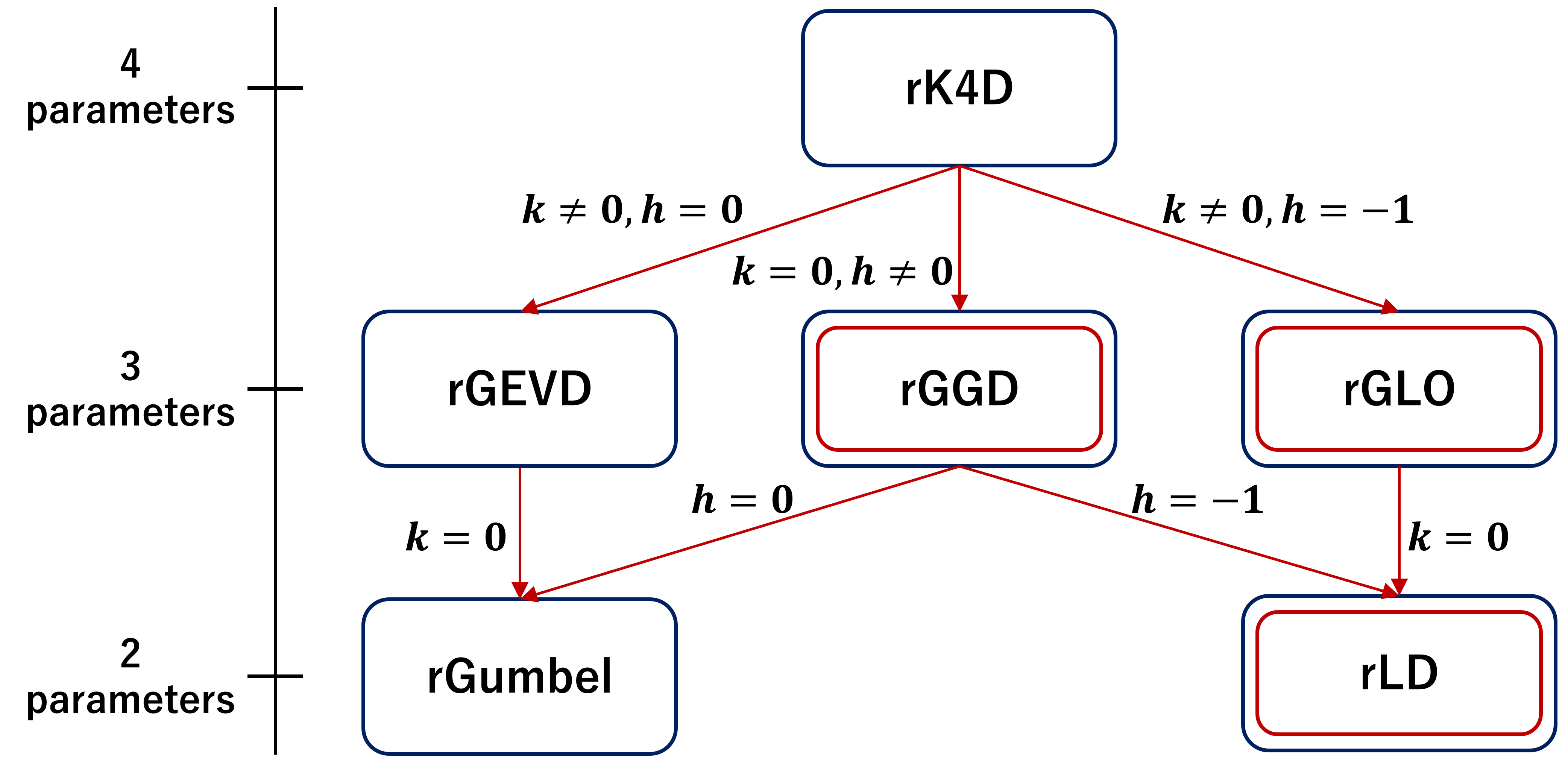} \end{tabular}	
	\caption{Relationship of the four-parameter kappa distribution for $r$ largest order statistics (rK4D) to other $r$ largest distributions. rGGD, rGLO, and rLD  are new distributions.} \label{rk4d_rel}
\end{figure}

Shin and Park (2023) proposed a four-parameter kappa distribution (K4D) for rLOS, as an extension of the rGEVD. We refer to it as the rK4D in this study. Because the univariate ($r=1$) K4D has been applied to represent a wider range of skew-kurtosis pairings and has captured more variability of observations than the three-parameter distributions (Murshed et al.~2014; Blum et al.~2017; Kjeldsen et al.~2017; Papukdee et al.~2022; Ibrahim 2022), we also expect the rK4D to cover a wider range of $r$ largest extreme data.
The rK4D includes some new distributions for rLOS as special cases, as illustrated  in Figure~ \ref{rk4d_rel}. New distributions are the generalized logistic distribution (GLO) for rLOS (rGLO), the logistic distribution for rLOS (rLD), and the generalized Gumbel distribution for rLOS (rGGD).

The GEVD sometimes yields inadequate results, especially for small to moderate sample sizes (Vogel and Wilson 1996; Salinas et al.~2014; Stein 2017). Thus, other many distributions, including the Pearson type III distribution or the GLO, have been considered to model extreme events (Hosking and Wallis 1997; Stein 2021a; Saulo et al.~2023, for example). The GLO was introduced by Ahmad et al.~(1988) as the log-logistic distribution and was reparametrized as the current version by Hosking and Wallis (1997). The GLO has often been importantly employed in researches for modeling hydrologic events (e.g., Kjeldsen et al.~2004; Fitzgerald 2005; Meshgi and Khalili 2008; Shin et al.~2010; Kim et al.~2015; Hussain et al.~2018) 
 Research into the choice of regional frequency distributions for flood data in the UK reported that the GLO was most often found to provide the best fit (Kjeldsen et al.~2017). Similar research for the maximum rainfall data in Malaysia found that the GLO is the best fit (Zin et al.~2009). {Based on our fitting to streamflow peaks over 424 stations in UK, the best $r$-LOS models selected are the rGEV at 162 (38.4$\%$) stations, the rGLO at 58 (13.6$\%$), the rLD at 30 (9.4$\%$), and other models at 164 (38.6$\%$) stations, in average over $r=1, 2, 3$. Thus we believe that the rGLO is still neccessary and useful, even though the rGEV or other models may be more frequently applied.}
 
In this study, we focus on the rGLO. {Even though the rGLO is a particular case of the rK4D, studying the rGLO more precisely should be meaningful because the rGLO is a generalization of the GLO. We expect and want to check that the rGLO provides the estimates with less standard error that the GLO does, because the rGLO ($r \ge 2$) incorporates more of the observed extreme data than the GLO ($r=1$).}  

Section~2 introduces the rK4D and provides some formulas for the rGLO. Section~3 discusses inferences, including the maximum likelihood estimation, delta method, profile likelihood approach, order selection by the entropy difference test, cross-validated likelihood criteria for model selection, and model averaging over $r$. Sections~4 and 5 illustrate the usefulness and practical effectiveness of rGLO using the Monte Carlo simulation and application to extreme flow data in Bevern Stream, UK. Section 6 presents the discussion, followed by the conclusion in Section 7. The Supplementary Material provides some details including formulas for the rLD and figures.

\section{The GLO for $r$ largest order statistics}
This section derives the density functions of the rGLO after introducing the rGEVD and rK4D. 

\subsection{rGEVD and rK4D} 
We denote $\underline{x}^r = (x^{(1)},x^{(2)},\cdots,x^{(r)})$ as the rLOS, among $n$ number of independent and identically distributed random variables following any distribution, satisfying $x^{(1)}\ge x^{(2)}\ge \cdots \ge x^{(r)}$. The GEV distribution for rLOS has the following joint density function (Coles 2001):
\begin{linenomath*}\begin{equation} \label{rgevd}
f_3(x^{(1)},x^{(2)}, \cdots,x^{(r)}) \;= \; \sigma^{-r} \text{exp} \left\{ - w(x^{(r)})^{1/k} \right\} \;
    \prod_{s=1}^r w(x^{(s)}) ^{{\frac{1}{k}} -1} ,
\end{equation}\end{linenomath*}
where $\sigma>0$, $ w(x^{(s)})= 1 -k \;{\frac{x^{(s)}-\mu}{\sigma}} >0$ for $s=1,2,\cdots,r$, and $\mu,\; \sigma$, and $k$ are the location, scale, and shape parameters, respectively.
The subscript in Eq.~(\ref{rgevd}) is used to specify three parameters to differentiate from the four-parameter distribution. The sign of $k$ in the above equations is changed from the book by Coles (2001). 
The particular case for $k=0$ in rGEVD is the rGD.

The PDF in Eq.~(\ref{rgevd}) is different from that of the rLOS of a random sample of the GEV distribution, but it is the PDF of the limiting joint distribution of the suitably rescaled rLOS from any distribution (Smith 1986).

As a generalization of some common three-parameter distributions including the GEVD, the K4D was introduced by Hosking (1994). The PDF of the K4D for $-\infty < \mu < \infty$,  $\sigma>0$, $k\neq0$, and $h\neq0$ is
\begin{linenomath*}\begin{equation} \label{k4d}
	f_4(x) \;=\; \sigma^{-1}w(x)^{(1/k)-1}\; F_4(x)^{1-h}  ,
	\end{equation}\end{linenomath*}
where
\begin{linenomath*}\begin{equation} \label{w(x)}
	w(x) \;=\; 1 -k\; {\frac{x-\mu}{\sigma}},
	\end{equation}\end{linenomath*} and
\begin{linenomath*}\begin{equation} \label{cdf4}
	F_4(x) \;=\;  {\left\{1-h\; {w(x)}^{1/k} \right\}}^{1/h}
	\end{equation}\end{linenomath*}
is the cumulative distribution function (CDF) of the K4D (Hosking 1994). Note that a new shape parameter $h$ is added from the GEVD. The K4D can be useful to represent a wider range of skew-kurtosis pairings and capture more variability of observations than the three-parameter distributions (Murshed et al.~2014).

Shin and Park (2023) proposed the rK4D, an extension of the rGEVD and the K4D. The rK4D is an analogous extension from the rGEVD by following a similar method as the extension from the GEVD to the K4D. The joint PDF of the rK4D under $\sigma>0$, $k\neq0$, and $h\neq0$ is
\begin{linenomath*}\begin{equation} \label{rk4d}
	f_4 (\underline{x}^r) \;=\; \sigma^{-r} C_r \times g(\underline{x}^r) \times F_4 (x^{(r)} )^{1-rh}.
	\end{equation}\end{linenomath*}
where
\begin{linenomath*} \begin{equation}  \label{C_r}
	C_r = 	\left\{ \begin{array}{ll} & \prod_{i=1}^{r-1}\; [1-(r-i)h], ~~~\textrm{if} ~ r\ge 2,  \\
	& 1, ~~~~~~~~~~~~~~~~~~~~~~~~~~~ \textrm{if}~ r=1, \end{array} \right. 	
	\end{equation}\end{linenomath*}	
$F_4$ is the CDF of K4D, as indicated in Eq.~(\ref{cdf4}), and
$	g(\underline{x}^r) \;=\; \prod_{s=1}^r w(x^{(s)})^{{\frac{1}{k}} -1} $,
where $\; w(x^{(s)}) \;=\; 1 -k {\frac{x^{(s)}-\mu}{\sigma}}\;$, as in Eq.~(\ref{w(x)}). The support for this PDF is $x^{(1)}\ge x^{(2)} \ge \cdots \ge x^{(r)}$, $ w(x^{(s)}) >0$ for $s=1,2,\cdots,r$, $C_r > 0$, and $\; 0< 1-h \; w(x^{(r)})^{1/k} <1$. From the requirement that $C_r >0$ in Eq.~(\ref{C_r}), we have the condition that $h < {\frac{1}{r-1}}$ for $r \ge 2$. When $r=1$, the PDF in Eq.~(\ref{rk4d}) is the same as the PDF of K4D in Eq.~(\ref{k4d}). When $h \rightarrow 0$, this PDF becomes the PDF of the rGEVD in Eq.~(\ref{rgevd}).
See Shin and Park (2023) for more details of the rK4D.

\subsection{The rGLO }
Following Hosking and Wallis (1997), the PDF of the GLO is
\begin{linenomath*}\begin{equation} \label{GLO}
	f^{\textit{\normalfont GL}}_3(x)\; =\; \sigma^{-1} \;{ \left\{ 1-k\frac{x-\mu}{\sigma} \right\}^{\frac{1-k}{k}} } \times [F^{\textit{\normalfont GL}}_3(x)] ^2 ,
	\end{equation}\end{linenomath*}
 for $-\infty < \mu < \infty$,  $\sigma>0$, and $ k\neq0$, where 
 \begin{linenomath*}\begin{equation} \label{GLO_cdf}
 	F^{\textit{\normalfont GL}}_3(x) \;=\; \frac{1}{1+\left\{ 1-k {\frac{x-\mu}{\sigma}}\right\}^{\frac{1}{k}}} \color{blue}{\ = \; 
 	\frac{1}{1+ w(x)^{\frac{1}{k}}} }
 	\end{equation}\end{linenomath*}
 is the CDF of the GLO. The superscript GL specifies the GLO. The quantiles of the GLO are
$\; z^{\text{GL}}_p \;=\; \mu +\dfrac{\sigma}{k} \left[ 1-\; \left(-\dfrac{p}{1-p}\right)^k \right]$,
where $\;F^{\text{GL}}_3 ( z^{\text{GL}}_p )= 1-p$.

Because the GLO is a special case ($h=-1$) of the K4D, we define the rGLO as a special case of the rK4D. The joint PDF of the rGLO is deduced from Eq.~(\ref{rk4d}) when $h=-1$, under $\sigma>0$, $k\neq0$,
\begin{linenomath*}\begin{equation} \label{rGLO}
f^{\textit{\normalfont GL}}_3 (\underline{x}^r) \;=\; \sigma^{-r} C^{\textit{\normalfont GL}}_r \times g^{\textit{\normalfont GL}}(\underline{x}^r) \times F_3^{\textit{\normalfont GL}} (x^{(r)} )^{(1+r)},
\end{equation}\end{linenomath*}
where
\begin{linenomath*} \begin{equation}  \label{GLO_C_r}
	C^{\textit{\normalfont GL}}_r = 	\left\{ \begin{array}{ll} & \prod_{i=1}^{r-1}\; [1+(r-i)], ~~~\textrm{if} ~ r\ge 2,  \\
	& 1, ~~~~~~~~~~~~~~~~~~~~~~~~~ \textrm{if}~ r=1, \end{array} \right. 	
	\end{equation}\end{linenomath*}	
 and
\begin{linenomath*}\begin{equation} \label{GLO_g(x)}
g^{\textit{\normalfont GL}}(\underline{x}^r) \;=\; \prod_{s=1}^r \;w(x^{(s)})^{{\frac{1}{k}} -1} ,
\end{equation}\end{linenomath*}
where $w(x)$ is defined in Eq.~(\ref{w(x)}). The support for this PDF is $x^{(1)}\ge x^{(2)}\ge \cdots \ge x^{(r)}$, $C^{\text{GL}}_r > 0$, and
\begin{linenomath*} \begin{equation} \begin{aligned}
& -\infty < x \le \mu+\sigma/k  ~~~\textrm{if} ~ k>0, \\
& \mu+\sigma/k \le x < \infty ~~~~~~~\textrm{if} ~ k<0.
\end{aligned}
 \end{equation}\end{linenomath*}	
When $r=1$, the PDF in Eq.~(\ref{rGLO}) is the same as the PDF of the GLO in Eq.~(\ref{GLO}).

The marginal PDF of the $s$th order statistic from the rGLO is derived using consecutive integrals of $f^{\textit{\normalfont GL}}_3 (\underline{x}^s)$ with respect to $(x^{(1)}, \cdots, x^{(s-1)})$ for $ 2 \le s \le r$:
\begin{linenomath*}\begin{equation} \label{marg_GLO} \begin{aligned}
   \color{blue}{ p^{\textit{\normalfont GL}}_{3}(x^{(s)}) } &\; =\; \int_{x^{(s)}}^{\infty}\int_{x^{(s-1)}}^{\infty}\cdots\int_{x^{(2)}}^{\infty}f^{\textit{\normalfont GL}}_{3}(\underline x^{s}) \;dx^{(1)}dx^{(2)} \dots dx^{(s-1)}\\
    &\;  =\; \frac{s}{\sigma}\times w(x^{(s)}) ^{\frac{s}{k}-1} \times \{ 1+   w(x^{(s)})
    ^{\frac{1}{k} } \}^{-(s+1)} \\
    &\color{blue}{\; = \; \frac{s}{\sigma}\times w(x^{(s)}) ^{\frac{s}{k}-1} \times  	F^{\textit{\normalfont GL}}_3(x^{(s)})^{s+1}  .}
      \end{aligned}
\end{equation}\end{linenomath*}	
\textcolor{blue}{Here, we use the notation $ p^{\textit{\normalfont GL}}_{3}(x^{(s)})$ to differentiate with $f^{\textit{\normalfont GL}}_3(x)$ in Eq.~(\ref{GLO}). }
The marginal CDF of the $s$th order statistic from the rGLO is obtained by integrating \textcolor{blue}{$p^{\textit{\normalfont GL}}_3 (x^{(s)})$}:
\begin{linenomath*}\begin{equation} \label{margcdf_GLO}
\begin{aligned}
	Pr(X^{(s)}<t) & \; \triangleq\; H^{\textit{\normalfont GL},(s)}_{3} ( t) \;=\; \int_{-\infty}^t \color{blue}{p^{\textit{\normalfont GL}}_3 (x^{(s)}) }\; dx^{(s)} \\
&\;=\; \int_{-\infty}^{t}\frac{s}{\sigma}\times  w(x^{(s)}) ^{\frac{s}{k}-1} \times \left\{1+ w(x^{(s)}) ^{\frac{1}{k}} \right\}^{-(s+1)}dx^{(s)}\\
	& 
	\;\color{blue}{ = \; 1- \left[1- 	F^{\textit{\normalfont GL}}_3(x^{(s)}) \right] ^{s} 
		.}
	\end{aligned}
	\end{equation}\end{linenomath*}	
The $(1-p)$ quantiles of the $s$th order statistic in the rGLO are obtained by inverting Eq.~(\ref{margcdf_GLO}): 
\begin{linenomath*}\begin{equation} \label{margzp_GLO}
z^{\text{GL},(s)}_p ~=~ \mu + {\frac{\sigma}{k}} \left[1- \left( \frac{p^{1/s}}{ 1-p^{1/s} } \right)^k \right],
	\end{equation}\end{linenomath*}	
for $s=1,2,\cdots, r$, where $\;H^{\text{GL},(s)}_3 ( z^{\text{GL},(s)}_p )= 1-p$.

The conditional PDF of $X^{(s)}$, given $\underline{X}^{s-1} = \underline{x}^{s-1}$
for $2 \le s \le r$ is 
\begin{linenomath*}\begin{equation} \label{condpdf_GLO} \begin{aligned}
f^{\textit{\normalfont GL}}_3 (x^{(s)} | \underline{x}^{s-1}) 
   \;=\;   {\frac{s}{\sigma}} \times g^{\textit{\normalfont GL}}_3(x^{(s)})\times \color{blue}{ \frac{
   F^{\textit{\normalfont GL}}_3 (x^{(s)})^{1+s}}{ F^{\textit{\normalfont GL}}_3(x^{(s-1)})^{s} } , } \\
   \end{aligned}
 	\end{equation}\end{linenomath*}	
 \textcolor{blue}{where $ g^{\textit{\normalfont GL}}_3(x^{(s)}) = w(x^{(s)})^{{\frac{1}{k}} -1}$. 
 This conditional PDF is the same as 
 \begin{equation}
 f^{\textit{\normalfont GL}}_3 (x^{(s)} | {x}^{(s-1)}) \;=\; \frac{ s\times f^{\textit{\normalfont GL}}_3 (x^{(s)})\; F^{\textit{\normalfont GL}}_3 (x^{(s)})^{(s-1)} } { F^{\textit{\normalfont GL}}_3 (x^{(s-1)})^s }, \nonumber
 \end{equation} 
 where $f^{\textit{\normalfont GL}}_3 (x^{(s)})$ is the PDF of the GLO in Eq.~(\ref{GLO}),} under $ x^{(s)} \le x^{(s-1)}$. Thus, the Markov property is satisfied similarly to the rGEVD and rK4D.

 The conditional CDF of $X^{(s)}$, given $\underline{X}^{s-1}$, is
 \begin{linenomath*}\begin{equation} \label{condcdfk4d}
 	H^{\textit{\normalfont GL}}_3 (x^{(s)} | \underline{x}^{s-1}) \;=\;   
 	\left(\frac{F^{\textit{\normalfont GL}}_3 (x^{(s)})} { F^{\textit{\normalfont GL}}_3
 		(x^{(s-1)})}\right)^{s} ,
 	\end{equation}\end{linenomath*}	
 under $ x^{(s)} \le x^{(s-1)}$. This result is the same as \textcolor{blue}{the $s$ power of} the unconditional CDF of the GLO with the right truncated at $x^{(s-1)}$ (Johnson et al.~1995). This property is used in generating random numbers from the rGLO.

\section{Inference methods for the rGLO }
\subsection{Maximum likelihood estimation }
We define  $\underline{x}^r_i =(x^{(1)}_i, x^{(2)}_i,\cdots,x^{(r)}_i)$, which is the $i$th  observation of the rLOS for $i=1,2,\cdots,m$, where $m$ is the sample size. By assuming $\{\underline{x}^r_1, \underline{x}^r_2, \cdots, \underline{x}^r_m \}$ follows the rGLO, the likelihood function of ($\mu, \sigma, k$) for $\sigma>0$, $k\neq0$, under constraints, is as follows:  
\begin{linenomath*}\begin{equation} \label{lf_GLO}
	L^{\textit{\normalfont GL}}(\mu, \sigma, k | \underline{x}^r )\; =\; \prod_{i=1}^m \;\left[ \sigma^{-r} C^{\textit{\normalfont GL}}_r \;[F^{\textit{\normalfont GL}}_3 (x^{(r)}_i)]^{(1+r)} \;
	\prod_{j=1}^r \left(1 - k\frac{x^{(j)}_i -\mu}{\sigma} \right)^{\frac{1}{k}-1} \right].
\end{equation}\end{linenomath*}	
The constraints that should be specified in numerically minimizing the negative log-likelihood function are $\sigma >0$, $ -\infty < x \le \mu+\sigma/k \;$ if $\; k>0$, and  $\; \mu+\sigma/k \le x < \infty $ if $\; k<0$.

\subsection{Delta method for variance estimation }
The $1/p$ return level ($z_p$) is defined as the $1-p$ quantile of the distribution (Coles 2001; Banfi et al.~2022). In most situations, researchers are interested in accurately estimating the return level for the first-order statistic. The return levels for the $s$th ($s> 1$) order statistics are less interesting. Thus, we consider mainly estimating the $1/p$ return level for the first-order statistic. For the annual extreme data, sometimes the name `T-year return level' is also used, with $T=1/p$. 

The variance estimation of the T-year return level can be calculated using the delta method, as follows (Coles 2001; Wang et al.~2017): 
$\;	\text{Var} (\hat z_p) \;\approx\; \nabla z_p^t \;V_r \;\nabla z_p $,
$V_r$ represents the covariance matrix of the parameter estimates, and
\begin{linenomath*}\begin{equation}\label{z_pGLO1}
\nabla z_p^t \;=\; [\dfrac{\partial z_{p}}{\partial\mu},\:\dfrac{\partial z_{p}}{\partial\sigma},\:\dfrac{\partial z_{p}}{\partial k}],
\end{equation}\end{linenomath*}
 evaluated at $(\hat \mu,\; \hat \sigma,\; \hat k)$, where $(\hat \mu,\; \hat \sigma,\; \hat k)$ indicate the maximum likelihood estimation (MLE) estimated from the rGLO. $V_r$ is usually approximated by the inverse of the observed Fisher information matrix.
The components in Eq.~(\ref{z_pGLO1}) for rGLO are
\begin{linenomath*}\begin{equation}\label{z_pGLO2}
\dfrac{\partial z^{\text{GL}}_p}{\partial\mu}=1, ~~~~
\dfrac{\partial z^{\text{GL}}_p}{\partial\sigma} \;= \;\dfrac{1-\left\{y^{-1}_p -1 \right\}^{k}}{k},
\end{equation}\end{linenomath*}
\begin{linenomath*}\begin{equation}\label{z_pGLO4}
\dfrac{\partial z^{\text{GL}}_p}{\partial k}\; =\; -\dfrac{\sigma}{k^{2}}\left[ -\left\{ y^{-1}_p-1\right\}^{k} \left\{ k\;\text{log}(y^{-1}_p -1)-1\right\}-1\right],
\end{equation}\end{linenomath*}
where $y_p=1-p$. 

\subsection{Profile likelihood for confidence interval }
The confidence interval of the return level based on the profile likelihood is more accurate than the symmetric interval based on the approximate normality (Coles 2001; Wang et al.~2017). This confidence interval for the return level of the first maxima based on the profile likelihood requires a reparameterization of the rGLO so that $z^{\text{GL},(1)}_p$ is one of the model parameters.
 Reparameterization is straightforward for the rGLO from Eq.~(\ref{margzp_GLO}) for $s=1,2,\cdots,r$:
\begin{linenomath*}\begin{equation} \label{profile_GLO}
	\mu^{\text{GL}}\;=\; z^{\text{GL},(s)}_p -\dfrac{\sigma}{k} \left[ 1- \left(\dfrac{p^{1/s}}{1-p^{1/s}}\right)^k \right],
\end{equation}\end{linenomath*}	
for a given $p$. Thus, the replacement of $\mu^{\text{GL}}$ in Eq.~(\ref{lf_GLO}) with Eq.~(\ref{profile_GLO}) has the desired effect of expressing the rGLO in terms of the parameters ($z_p^{\text{GL},(1)}$, $\sigma$, $k$). Afterward, the profile log-likelihood is obtained by maximization with respect to the remaining parameters in the usual way. The $95\%$ confidence interval for $z_p$ is obtained by calculating a value at a height of 0.5 $\times$ $\chi_1^2(.05)$ below the maximization of profile log-likelihood, where $\chi_1^2(.05)$ is the .95 quantile of the $\chi_1^2$ distribution, and determining the points of intersection (Coles 2001).

\subsection{Selection of $r$ using the entropy difference test}
	
	 Using the $r$ largest extremes enhances the estimation power for moderate values of $r$. The number $r$ comprises a bias-variance trade-off. Small values of $r$ consist of few data points leading to high variance, whereas large values of $r$ are likely to violate the asymptotic support for the model, leading to bias (Coles 2001). 
	Bader et al.~(2017) and Silva et al.~(2021) developed automated methods for selecting $r$ from the rGEVD. 
	The rationale of choosing a larger value of $r$ is to use much information as possible, but not set $r$ too high so that the rGEV approximation becomes poor due to the decrease in convergence rate (Bader et al.~2017).
	
{Bader et al.~(2017) proposed two test methods to select an appropriate $r$ in the rGEV model. The first one is a score test, and the second one is the entropy difference test. In this study, we employ the second one with a modification for the rGLO, in which the difference in estimated entropy between $\text{GLO}_{r}$ and $\text{GLO}_{r-1}$ is used. 
	The hypothesis for this test are $H_0 :~ r^* =r $ vs. $H_1 :~ r^* = r-1 $ for $r=2,\cdots, R$, where $r^* $ is the best appropriate top order, and $R$ is the maximum, predetermined number of top order.}

{The entropy for a continuous random variable with density $f$ is defined by
	\begin{linenomath*}\begin{equation}\label{entropy}
		E[-\text{ln} f(y)] = - \int_{-\infty}^{\infty} f(y) \text{log} f(y) dy.
		\end{equation}\end{linenomath*}
	It is essentially the expectation of negative log-likelihood. The expectation can be approximated with the sample average of the contribution to the negative log-likelihood from the observed data, or simply the log-likelihood scaled by the sample size $n$. The difference in the negative log-likelihood between $\text{GLO}_{r-1}$ and $\text{GLO}_{r}$ provides a measure of deviation from $H_{0}$. Large deviation from the expected difference under $H_{0}$ suggests a possible misspecification of $H_{0}$ (Bader et al.~2017).}
	
{From the log-likelihood contribution in Eq.~(\ref{lf_GLO}), the difference in negative log-likelihood for the $i$th block, $Y_{ir} = l^{(r)}_{i}-l^{(r-1)}_{i}$, is
	\begin{linenomath*}\begin{equation}\label{entropy}
		\begin{aligned}
		Y_{ir} = -\text{log}\sigma + & \text{log}(r) 
		- r\text{log}\left[{1+\left(1-k\frac{x_i^{(r-1)}-\mu}{\sigma}\right)^{1/k}} \right] \\
		& + (1+r)\text{log}\left[{1+\left(1-k\frac{x_i^{(r)}-\mu}{\sigma}\right)^{1/k}} \right] 
		+(\frac{1}{k}-1)\text{log}\left(1-k\frac{x_i^{(r)}-\mu}{\sigma} \right)
		\end{aligned}
		\end{equation}\end{linenomath*}
	Let $\bar{Y}_{r}=\frac{1}{n}\sum_{i=1}^n Y_{ir}$. Bader et al.~(2017) considered a standardized version of $\bar{Y}_{r}$:
	\begin{equation}
	T_{rn} =\sqrt(n)(\bar{Y}_{r} - \eta_r)/ S_{y_r},
	\end{equation} 
	where $\eta_r$ is the expected value of $\bar{Y}_{r}$, and $S_{y_r} = \frac{1}{n-1}\sum_{i=1}^n (Y_{ir}-\bar{Y}_{r})^2$. Then $T_{rn}$ converges in distribution to the standard normal, under the null hypothesis. We used this method for the rGLO model.}

 \subsection{Model averaging over $r$ for quantile estimation}

 In addition to choosing the best $r$, we consider a model averaging (or ensemble) approach in this study. An ensemble estimate of $1/p$ return level is defined as a weighted sum;
 \begin{equation}
 z_p^E = w_1 z_p^{(1)} + w_2 z_p^{(2)} + \cdots + w_r z_p ^{(r)},
 \end{equation}
 where $z_p^{(j)}$ is $1/p$ return level estimated from the $GLO_{(j)}$ for $j=1,2,...,r$, and $w_j$ is the weight for the $GLO_{(j)}$ . 
 Model averaging approach has been shown to have a good prediction performance and robust against model misspecification. Model averaging is a means of allowing for model uncertainty in estimation which can provide better estimates and more reliable confidence intervals than model selection (Fletcher, 2018; Okoli et al.~2018; Salaki et al.~2022; Galavi et al.~2023, for example). 

 For a weighting scheme, we considered $ \text{Var} (\hat z_p^{(j)})$ which is available by the delta method. To give bigger weight to a model with smaller value of the variance estimate, we used the following weight for the $GLO_{(j)}$ model for $j=1,2,...,r$;
 \beq \label{weight}
 w_j \;=\;{ \frac{1}{\text{Var} (\hat z_p^{(j)})} \over {\sum_{j=1}^r \frac{1}{\text{Var} (\hat z_p^{(j)})} } }.
 \eeq
 We expect this model averaging provides a robust performance than a rGLO model with the selected $r$.

 We considered another weighting scheme based on the generalized L-moments distance, following Yoon at al.~(2023). But the performance measures obtained from Monte Carlo simulation did not show better result than the above variance-based weighting scheme. The details of this approach are described in the Supplementary Material. However, this weighting scheme based on the generalized L-moments distance is applied to a real data study in Section 5.
 
 \subsection{Cross-validated likelihood approach for model selection}
 
{Cross-validated (CV) likelihood was investigated by Smyth (2000) as a tool for determining the appropriate number of components in finite mixture modeling. The CV likelihood is defined as computed the likelihood for the test data into which the parameter estimates (such as the MLE) obtained from the training data are pluged. The CV log-likelihood (CVLL) is an unbiased estimator of the Kullback-Leibler distance between truth and the models under consideration, and this motivates to use the CVLL as a model selection criterion (Smyth 2000). The CVLL has been extensively used in many purposes in statistical research.
 	For example, Matsuda et al.~(2006) and Stein (2021b) employed the CVLL to compare models or select a good model. van Wieringen and Chen (2021) and van Wieringen and Binder (2022) applied the k-fold CVLL in penalized estimations, and Fauer and Rust (2023) used k-fold CV BIC for non-stationary extreme value model selection. 	
 	In this study, we also compute the CVLL, CV AIC, and CV BIC for model comparison, based on 5-fold cross-validation.} 	
 
\section{Monte Carlo simulation}
\subsection{Simulation setting}
To understand the estimation performance of the rGLO and rLD, we executed a simulation study where some high quantiles are already known, considering two cases. The first is generating random numbers from the rGLO, a simulation using a known population. The second case is generating random numbers from a distribution different from the rGLO. For such a different distribution, the rK4D is considered in this study. The latter approach is called as a simulation using an ``unknown population”. 

This paper focuses on estimating the 100-year return level (.99 quantile) of the first-order statistic. We generated 1,000 random samples to calculate the following evaluation measures to compare several estimators:
\begin{linenomath*}\begin{equation} \label{eval}
	\begin{aligned}
&\text{Bias}\; = \;{\frac{1}{M}} \sum_{i=1}^M (\hat q_i - q) \;= \;\bar {\hat {q}} - q ,\\
&\text{SE} \;= \;\sqrt{ \text{Var}(\hat q_i) }, \\
&\text{RMSE} \;= \;\sqrt{  {\frac{1}{M}} \sum_{i=1}^M (\hat q_i - q) ^2 },
   \end{aligned}
\end{equation}\end{linenomath*}	
where $\hat{q}_{i}$ and $q$ are the estimated and true quantiles of the first maximum, respectively; $\bar{\hat{q}}= \frac{1}{M}\sum^{M}_{i=1} \hat{q}_{i}$, and $M$ represents the number of successful convergences among 1,000 trials. $\text{SE}$ and $\text{RMSE}$ denote the standard error and root mean squared error of the estimator. Smaller in absolute is better.  

\subsection{Random number generation from the rGLO }
The property that Eq.~(\ref{condcdfk4d}) is \textcolor{blue}{the $s$ power of} the CDF of the GLO with the right truncated at $x^{(s-1)}$ is exploited to generate the $r$ components in a realized rGLO observation, as Bader et al.~(2017) did for the rGEVD. The pseudo algorithm to generate a single observation from the rGLO follows:
\begin{itemize}
\item[1.] Generate $U_1, \dots, U_r$, where $U$ values are random numbers from the uniform (0,1) distribution.
\textcolor{blue}{	\item[2.] Truncate uniform random numbers ($W_1, \dots, W_r$) as $\ W_s = \prod_{j=1}^s\ (U_j)^{b_j} \ $ for $s=1,\dots,r$, where $\ b_j = 1 /j $.}
\item[3.] Obtain $x^{(i)} = F^{-1} (W_i)$, where $F$ is the CDF of the GLO, as given in Eq.~(\ref{GLO_cdf}).
\end{itemize}
\noindent The resulting vector $(x^{(1)}, \dots, x^{(r)} )$ is a single observation from the rGLO.

\begin{table}[htb!]
	\centering
	\caption{Result of Monte Carlo experiments: bias, standard error (SE), and root mean squared error (RMSE) for estimating the 100-year return level obtained from the rGLO models with the MLE method under shape parameter $k$ varying from -.3 to .3 as $r$ changes from 1 to 6. The true values and estimates of 100-year return level are also provided.}
	\label{shin_table}
	\vspace{0.5 cm}
	\begin{tabular}{c|c|cccccccc}
		\hline
		\multirow{2}{*}{measure}    & \multirow{2}{*}{r}       & \multicolumn{8}{c}{k}                                                                                                                 \\ \cline{3-10}
		&                          & -0.3           & -0.2           & -0.1           & -0.05          & 0.05           & 0.1            & 0.2            & 0.3            \\ \hline
		\multirow{6}{*}{Bias}      & 1                        & -5.48          & -3.58          & -2.20          & -1.71          & -1.11          & -0.87          & -0.56          & -0.32          \\
		& 2                        & -2.07          & -1.12          & -0.58          & -0.37          & -0.18          & \textbf{-0.12} & \textbf{-0.06} & \textbf{0.01}  \\
		& 3                        & -1.67          & -0.51          & \textbf{-0.01} & \textbf{0.13}  & \textbf{0.18}  & 0.17           & 0.12           & 0.11           \\
		& 4                        & \textbf{-1.60} & \textbf{-0.12} & 0.36           & 0.44           & 0.37           & 0.30           & 0.16           & 0.11           \\
		& 5                        & -1.66          & 0.16           & 0.62           & 0.64           & 0.44           & 0.33           & 0.15           & 0.09           \\
		& 6                        & -1.93          & 0.35           & 0.79           & 0.75           & 0.45           & 0.32           & 0.11           & 0.07           \\ \hline
		\multirow{6}{*}{SE}        & 1                        & 6.05           & 3.96           & 2.43           & 1.94           & 1.21           & 0.97           & 0.61           & 0.42           \\
		& 2                        & 3.63           & 2.33           & 1.52           & 1.19           & 0.80           & 0.64           & 0.40           & 0.26           \\
		& 3                        & 3.19           & 1.92           & 1.20           & 0.94           & 0.64           & 0.53           & \textbf{0.33}  & \textbf{0.23}  \\
		& 4                        & 3.14           & 1.78           & 1.08           & 0.85           & \textbf{0.61}  & \textbf{0.51}  & 0.33           & 0.24           \\
		& 5                        & \textbf{2.96}  & 1.61           & 0.98           & 0.80           & 0.61           & 0.53           & 0.35           & 0.25           \\
		& 6                        & 3.00           & \textbf{1.53}  & \textbf{0.94}  & \textbf{0.79}  & 0.64           & 0.56           & 0.37           & 0.26           \\ \hline
		\multirow{6}{*}{RMSE}      & 1                        & 8.16           & 5.34           & 3.28           & 2.59           & 1.64           & 1.30           & 0.83           & 0.52           \\
		& 2                        & 4.18           & 2.59           & 1.62           & 1.25           & 0.82           & 0.66           & 0.40           & 0.26           \\
		& 3                        & 3.60           & 1.98           & 1.20           & \textbf{0.95}  & \textbf{0.66}  & \textbf{0.55}  & \textbf{0.35}  & \textbf{0.25}  \\
		& 4                        & 3.52           & 1.78           & \textbf{1.14}  & 0.96           & 0.71           & 0.59           & 0.37           & 0.26           \\
		& 5                        & \textbf{3.39}  & 1.62           & 1.16           & 1.02           & 0.76           & 0.63           & 0.38           & 0.27           \\
		& 6                        & 3.57           & \textbf{1.57}  & 1.23           & 1.09           & 0.79           & 0.65           & 0.39           & 0.27           \\ \hline
		\multirow{7}{*}{100yr\_rl} & 1                        & 25.38          & 21.11          & 18.03          & 16.88          & 15.21          & 14.55          & 13.57          & 12.81          \\
		& 2                        & 21.97          & 18.65          & 16.41          & 15.54          & 14.29          & \textbf{13.80} & \textbf{13.07} & \textbf{12.48} \\
		& 3                        & 21.57          & 18.04          & \textbf{15.84} & \textbf{15.04} & \textbf{13.93} & 13.51          & 12.89          & 12.38          \\
		& 4                        & \textbf{21.50} & \textbf{17.65} & 15.47          & 14.73          & 13.74          & 13.38          & 12.85          & 12.38          \\
		& 5                        & 21.56          & 17.37          & 15.21          & 14.53          & 13.67          & 13.35          & 12.86          & 12.40          \\
		& 6                        & 21.83          & 17.18          & 15.04          & 14.42          & 13.66          & 13.36          & 12.90          & 12.42          \\ \hline
		& \multicolumn{1}{c|}{True} & 19.90          & 17.53          & 15.83          & 15.17          & 14.11          & 13.68          & 13.01          & 12.49 \\ \hline
	\end{tabular}
\end{table}

\subsection{Experiments with a known population}

Table~\ref{shin_table} presents the Monte Carlo experiment results from the rGLO with $\mu=0$, $\sigma=1$, and sample size $n=30,\; 60$. The bias, SE, and RMSE are presented for estimating the 100-year return level obtained from rGLO models and model averaging method under $k$ varying from -.2 to .2 and $r$ changing from 1 to 5. 
The true values and estimates of 100-year return level are also provided. The best performance is indicated in bold font.

\begin{table}[htb!]
	\caption{Same as Table \ref{shin_table} but for using model averaging method.}
	\label{MA_table}
	\centering
	\vspace{0.3cm}
	\begin{tabular}{c|c|cccccccc}
		\hline
		\multirow{2}{*}{measure}    & \multirow{2}{*}{r} & \multicolumn{8}{c}{k}                                                                                                                 \\ \cline{3-10}
		&                    & -0.3           & -0.2           & -0.1           & -0.05          & 0.05           & 0.1            & 0.2            & 0.3            \\ \hline
		\multirow{6}{*}{Bias}      & 1                  & -5.48          & -3.58          & -2.20          & -1.71          & -1.11          & -0.87          & -0.56          & -0.32          \\
		& 2                  & -1.82          & -1.06          & -0.61          & -0.45          & -0.32          & -0.27          & -0.21          & -0.11          \\
		& 3                  & -1.24          & -0.50          & -0.16          & \textbf{-0.07} & \textbf{-0.05} & \textbf{-0.05} & -0.07          & -0.02          \\
		& 4                  & \textbf{-1.11} & -0.21          & \textbf{0.11}  & 0.16           & 0.11           & 0.08           & \textbf{0.01}  & \textbf{0.02}  \\
		& 5                  & -1.14          & \textbf{0.00}  & 0.30           & 0.32           & 0.21           & 0.15           & 0.04           & 0.03           \\
		& 6                  & -1.27          & 0.15           & 0.45           & 0.44           & 0.26           & 0.18           & 0.06           & 0.04           \\ \hline
		\multirow{6}{*}{SE}        & 1                  & 6.05           & 3.96           & 2.43           & 1.94           & 1.21           & 0.97           & 0.61           & 0.42           \\
		& 2                  & 3.47           & 2.26           & 1.47           & 1.15           & 0.78           & 0.63           & 0.41           & 0.28           \\
		& 3                  & 2.99           & 1.83           & 1.16           & 0.93           & 0.65           & 0.54           & 0.35           & 0.25           \\
		& 4                  & 2.83           & 1.69           & 1.06           & 0.85           & 0.61           & 0.52           & 0.34           & 0.24           \\
		& 5                  & 2.71           & 1.57           & 0.99           & 0.81           & \textbf{0.60}  & \textbf{0.51}  & \textbf{0.34}  & \textbf{0.24}  \\
		& 6                  & \textbf{2.70}  & \textbf{1.51}  & \textbf{0.95}  & \textbf{0.79}  & 0.60           & 0.52           & 0.34           & 0.24           \\ \hline
		\multirow{6}{*}{RMSE}      & 1                  & 8.16           & 5.34           & 3.28           & 2.59           & 1.64           & 1.30           & 0.83           & 0.52           \\
		& 2                  & 3.92           & 2.50           & 1.59           & 1.24           & 0.84           & 0.69           & 0.46           & 0.30           \\
		& 3                  & 3.24           & 1.90           & 1.17           & 0.93           & 0.65           & 0.54           & 0.36           & 0.25           \\
		& 4                  & 3.04           & 1.70           & 1.06           & \textbf{0.86}  & \textbf{0.62}  & \textbf{0.52}  & \textbf{0.34}  & \textbf{0.24}  \\
		& 5                  & \textbf{2.94}  & 1.57           & \textbf{1.04}  & 0.87           & 0.63           & 0.53           & 0.34           & 0.24           \\
		& 6                  & 2.98           & \textbf{1.51}  & 1.05           & 0.90           & 0.66           & 0.55           & 0.34           & 0.25           \\ \hline
		\multirow{7}{*}{100yr\_rl} & 1                  & 25.38          & 21.11          & 18.03          & 16.88          & 15.21          & 14.55          & 13.57          & 12.81          \\
		& 2                  & 21.72          & 18.59          & 16.44          & 15.62          & 14.43          & 13.95          & 13.22          & 12.60          \\
		& 3                  & 21.14          & 18.03          & 15.99          & \textbf{15.24} & \textbf{14.16} & \textbf{13.73} & 13.08          & 12.51          \\
		& 4                  & \textbf{21.01} & 17.74          & \textbf{15.72} & 15.01          & 14.00          & 13.61          & \textbf{13.00} & \textbf{12.47} \\
		& 5                  & 21.04          & \textbf{17.53} & 15.53          & 14.85          & 13.90          & 13.54          & 12.97          & 12.46          \\
		& 6                  & 21.17          & 17.38          & 15.38          & 14.73          & 13.85          & 13.50          & 12.95          & 12.45          \\ \hline
		& True             & 19.90          & 17.53          & 15.83          & 15.17          & 14.11          & 13.68          & 13.01          & 12.49          \\ \hline
	\end{tabular}
\end{table}

  As $r$ moves from 1 to 6, the bias decreases for a while but increases at $r=3$ or 4 or 5. The SE also decreases at first but increases after $r=3$. Thus the similar pattern is observed for the RMSE. For $k \le -.1$, the RMSE has minimum at $r=4$ or 5 or 6. Whereas, for $k \ge -.05$, the RMSE has minimum at $r=3$. For each $r$, SE and RMSE decrease to relatively small values compared to the true value, as $k$ changes from -.3 to .3. Figure S1 in the Supplementary Material shows these patterns graphically. For each $k$, all values of bias, SE, RMSE for $r \ge 2$ are smaller than those values for $r=1$, which means that the rGLO model using $r (\ge 2)$ LOS is better than the GLO distribution using just block maxima ($r=1$).
  
  Biases at $r=1$ are not close to zero. One can raise a question why biases are not close to zero even though the GLO model was fitted to data generated from the GLO distribution. We think the reason is because estimating (or extrapolating) high quantiles based on small or moderate sample is inherently very difficult. Moreover, another reason is that the MLE may not work well for small or moderate sample, especially for negative $k$. This non-zero bias or poor performance of the MLE for negative $k$ has been also reported when the MLE is used for the GEV distribution (Hosking et al.~1985; Martins and Stedinger 2000; Yoon et al.~2023) or for the K4D (Papukdee et al.~2022).
    
   Bias and SE are expected to be decreased with more data as $r$ increases, which are observed in Table \ref{shin_table}. But, bias and SE do not decrease straightly as $r$ increases. The reason of non-straight decrease is because bias and SE are calculated for the quantile estimate of the first maximum (at $r=1$). The rGLO model covers not just the first maximum but covers all $r$ largest order statistics. Thus, as $r$ moves to higher values than $r=1$, the focus of estimation by the rGLO moves from the first maximum to higher order statistics. Therefore, especially when $r >> 1$, there are high possibility that the quantile estimate of the first maximum based on the rGLO diverges from the true quantile of the first maximum. 
  
 Table \ref{MA_table} presents the Monte Carlo experiment results from the model averaging approach with weights Eq.~(\ref{weight}) based on variance estimates of $1/p$ return level. Changing patterns of Bias, SE, and RMSE, as $r$ increases or $k$ changes, are similar to Table \ref{shin_table}. Figure S2 in the Supplementary Material shows this pattern. An important finding is that minimum values of RMSE obtained from the model averaging method in Table \ref{MA_table} are always smaller than those RMSE values from single rGLO model in Table \ref{shin_table}, for every $k$.

\begin{table}[h!]
	\caption{Result of three Monte Carlo experiments in the four-parameter kappa distribution for $r$ largest order statistics (rK4D) for parameters with $h=-.5,\; -.7,\; -.9,\; -1.1$ for $k=-.1$ and with sample size $n=30, 60$. Bias, standard error (SE), and root mean squared error (RMSE) calculated from three models are presented for the 100-year return levels as $r$ changes from 1 to 5.}
	\vspace{0.4cm}
	\resizebox{0.80\textwidth}{!}{\begin{minipage}{\textwidth}
			\label{tab_rk4d}
			\begin{tabular}{c|c|c|ccc|ccc|ccc|ccc}
				\hline
				\multirow{2}{*}{n}   & \multirow{2}{*}{Model} & \multirow{2}{*}{r} & \multicolumn{3}{c|}{$h=-0.5$}            & \multicolumn{3}{c|}{$h=-0.7$}              & \multicolumn{3}{c|}{$h=-0.9$}              & \multicolumn{3}{c}{$h=-1.1$}              \\ \cline{4-15}
				&                        &                    & BIAS          & SE             & RMSE           & BIAS          & SE             & RMSE           & BIAS          & SE             & RMSE           & BIAS          & SE             & RMSE           \\ \hline
				\multirow{15}{*}{30} & \multirow{5}{*}{rGLO}  & 1                  & -10.4         & 804.4          & 912.7          & -5.3          & 518.8          & 547.1          & -7.2          & 835.5          & 887.7          & -1.9          & 489.1          & 492.6          \\
				&                        & 2                  & -6.0          & 424.0          & 459.9          & \textbf{-0.4} & 334.6          & 334.7          & \textbf{-0.9} & 377.8          & 378.7          & \textbf{0.5}  & 348.1          & 348.4          \\
				&                        & 3                  & -3.4          & 391.1          & 402.4          & 1.5           & 268.4          & \textbf{270.6} & 2.6           & 268.3          & 274.9          & 2.8           & 272.7          & 280.5          \\
				&                        & 4                  & \textbf{1.3}  & 299.4          & 301.0          & 4.2           & 271.7          & 289.7          & 4.8           & 227.6          & 250.4          & 5.1           & 254.3          & 280.5          \\
				&                        & 5                  & 4.0           & \textbf{260.6} & \textbf{276.2} & 7.3           & \textbf{220.2} & 273.0          & 7.0           & \textbf{182.6} & \textbf{231.8} & 6.6           & \textbf{220.6} & \textbf{264.6} \\ \cline{2-15}
				& \multirow{5}{*}{rK4D}  & 1                  & \textbf{5.6}  & 443.7          & 475.5          & \textbf{9.6}  & 325.3          & 417.5          & \textbf{6.9}  & 369.4          & 417.5          & \textbf{8.9}  & 315.5          & 394.1          \\
				&                        & 2                  & 6.9           & 315.9          & 363.3          & 9.6           & 255.0          & 346.3          & 9.1           & 257.2          & 339.3          & 9.9           & 237.0          & \textbf{335.0} \\
				&                        & 3                  & 8.2           & 263.0          & 330.2          & 11.2          & 211.1          & \textbf{336.9} & 10.8          & 207.5          & 324.7          & 10.9          & 218.6          & 337.4          \\
				&                        & 4                  & 10.0          & 221.0          & 320.4          & 12.3          & 201.6          & 353.0          & 11.6          & 188.8          & 322.1          & 11.7          & 208.5          & 345.2          \\
				&                        & 5                  & 11.0          & \textbf{195.2} & \textbf{316.9} & 13.2          & \textbf{187.8} & 361.0          & 12.0          & \textbf{176.0} & \textbf{320.7} & 12.2          & \textbf{199.1} & 346.7          \\ \cline{2-15}
				& \multirow{5}{*}{rGEV}  & 1                  & \textbf{2.1}  & 569.9          & 574.5          & \textbf{5.7}  & 379.9          & 412.5          & \textbf{3.8}  & 445.3          & 459.9          & \textbf{6.2}  & 340.9          & 379.4          \\
				&                        & 2                  & 7.0           & 279.8          & 328.3          & 9.6           & 229.6          & 322.6          & 8.8           & 252.3          & 330.1          & 9.6           & 229.4          & \textbf{321.2} \\
				&                        & 3                  & 9.0           & 230.9          & \textbf{312.3} & 11.4          & 187.5          & \textbf{316.5} & 10.8          & 199.7          & \textbf{317.1} & 10.9          & 202.9          & 321.8          \\
				&                        & 4                  & 10.7          & 209.7          & 325.2          & 12.6          & 181.8          & 340.1          & 11.7          & 182.8          & 318.5          & 11.6          & 197.0          & 332.2          \\
				&                        & 5                  & 11.7          & \textbf{194.1} & 330.6          & 13.6          & \textbf{175.5} & 360.1          & 12.3          & \textbf{172.9} & 323.6          & 12.1          & \textbf{188.7} & 333.9          \\ \hline
				\multirow{15}{*}{60} & \multirow{5}{*}{rGLO}  & 1                  & -6.7          & 302.6          & 347.2          & -4.6          & 262.0          & 283.2          & \textbf{-0.5} & 204.7          & 205.0          & \textbf{-0.6} & 193.9          & 194.2          \\
				&                        & 2                  & -3.8          & 191.7          & 206.3          & \textbf{-0.6} & 183.0          & 183.3          & 1.9           & 158.6          & 162.3          & 1.0           & 144.5          & 145.4          \\
				&                        & 3                  & \textbf{-1.2} & 147.5          & 148.9          & 1.3           & 153.8          & 155.4          & 3.9           & 113.5          & \textbf{129.0} & 2.6           & 134.4          & \textbf{141.2} \\
				&                        & 4                  & 1.8           & 128.5 & \textbf{131.7} & 4.0           & 131.6          & \textbf{147.6} & 5.5           & 99.9           & 129.6          & 3.9           & 127.6          & 142.5          \\
				&                        & 5                  & 3.6           & \textbf{118.5}          & 131.7          & 5.8           & \textbf{120.0} & 153.2          & 7.2           & \textbf{86.6}  & 138.5          & 4.7           & \textbf{125.7} & 148.0          \\ \cline{2-15}
				& \multirow{5}{*}{rK4D}  & 1                  & \textbf{3.3}  & 213.5          & 224.2          & \textbf{4.9}  & 186.2          & 209.9          & \textbf{6.9}  & 143.4          & \textbf{191.6} & \textbf{5.7}  & 181.8          & 214.5          \\
				&                        & 2                  & 5.5           & 153.5          & 183.3          & 6.8           & 145.2          & \textbf{192.1} & 8.8           & 114.7          & 192.8          & 7.4           & 140.2          & \textbf{194.3} \\
				&                        & 3                  & 7.4           & 122.4          & \textbf{176.6} & 8.5           & 121.5          & 193.9          & 10.2          & 92.6           & 196.9          & 8.2           & 127.0          & 194.7          \\
				&                        & 4                  & 8.8           & 107.6          & 184.3          & 9.5           & 113.3          & 203.6          & 10.6          & 85.4           & 197.4          & 8.5           & \textbf{126.1} & 198.8          \\
				&                        & 5                  & 9.6           & \textbf{101.8} & 194.2          & 9.9           & \textbf{108.0} & 206.4          & 10.8          & \textbf{85.0}  & 202.3          & 8.5           & 128.0          & 200.3          \\ \cline{2-15}
				& \multirow{5}{*}{rGEV}  & 1                  & \textbf{3.3}  & 198.7          & 209.7          & \textbf{4.8}  & 178.8          & 201.9          & \textbf{7.4}  & 137.0          & \textbf{191.1} & \textbf{5.9}  & 163.8          & 198.1          \\
				&                        & 2                  & 6.5           & 135.7          & \textbf{177.4} & 7.8           & 130.5          & \textbf{191.0} & 9.4           & 107.9          & 196.1          & 7.6           & 134.8          & \textbf{191.8} \\
				&                        & 3                  & 8.4           & 111.2          & 181.7          & 9.1           & 116.7          & 200.4          & 10.4          & 94.9           & 203.8          & 8.4           & \textbf{130.6} & 201.0          \\
				&                        & 4                  & 9.6           & 104.5 & 196.5          & 10.0          & 112.7          & 212.6          & 10.9          & \textbf{91.1}  & 209.6          & 8.7           & 132.8          & 207.6          \\
				&                        & 5                  & 10.2          & \textbf{100.7}          & 204.5          & 10.4          & \textbf{111.7} & 219.8          & 11.1          & 91.2           & 215.3          & 8.7           & 135.4          & 211.2          \\ \hline
			\end{tabular}
	\end{minipage} }
\end{table}

\subsection{Experiments with an unknown population}

{For the fair performance evaluation of the considered distributions, the random samples from the rK4D were used as if they were generated from the true distribution, because the true underlying distribution is never known in practice. Random samples with with sample sizes $n=30,\;60$ were generated under setting of $h=-.5,\; -.7,\; -.9,\; -1.1$ for $k=-.2$.  The location and scale parameters, $\mu$ and $\sigma$, of the rK4D were set to 0 and 1, respectively (see the work by Shin and Park (2023) for details on generating random samples from the rK4D).
 Because the rK4D reduces to rGLO when the shape parameter is $h=-1$, the setting $h=-.9, \;-1.1$ can be regarded as generating slightly distorted samples from the relevant rGLO. The result may indicate how the rGLO is affected by the condition when the assumed distribution differs from the parent distribution and how much useful the rGLO model is in practice.}

{Table~\ref{tab_rk4d} summarizes this simulation for the 100-year return level estimation using the maximum likelihood method. The performance measures (i.e., bias, SE, and RMSE) for three models (rGLO, rGEVD, and rK4D) are presented as $r$ changes from 1 to 5. Figure~S3 in the Supplementary Material 
	graphically summarizes this simulation.}

{In Table~\ref{tab_rk4d} and Figure~S1, 
the SEs and RMSEs for all models generally decreased as $r$ changed from 1 to 5. Thus, the positive effect of using the rGLO is obtained when applying it to the rLOS ($r \ge 2$) instead of applying it to the block maxima only with the GLO ($r=1$).
 In comparison, more improvement of the RMSE in the rGLO is observed than in the rGEVD as $r$ changes from 1 to 5. The RMSEs of the rGLO are generally smaller than those of the rK4D.} 

{In this simulation, one may be curious why the rK4D did not work well even though random samples were generated from the rK4D. One reason may be found from the result (Shin and Park 2023) that the MLE for the rK4D did not work well. If we used the maximul ``penalized" likelihood estimation, the results for the rK4D have been changed to obtain better RMSEs.}

 \section{Real application: Extreme flow in Bevern Stream}
 For a real application, we consider the extreme flow data observed daily in Bevern Stream (station ID: 41020) in Clappers Bridge, UK. The data are available from the Peak Flow Dataset website, https://nrfa.ceh.ac.uk/peak-flow-dataset. The data in Bevern Stream, \textcolor{blue}{used for this study,} comprise the three largest (up to $r=3$) flows from 1969 to 2021 for 52 years (except 1973), in meters cubed per second. 
 The reason why we chose a station Bevern Stream is because the rGLO model was selected as the best model among some $r$-LOS models fitted to the data of Bevern Stream.

  Among 424 streamflow stations in UK where we have tried to fit various $r$-LOS models, the rGLO was selected as the best model at 58 stations, when a decision was made based on the Bayesian information criterion. Actually, the best $r$-LOS models selected for those streamflow stations in UK are the rGEV at 162 (38.4$\%$) stations, the rGLO at 58 (13.6$\%$), the rLD at 30 (9.4$\%$), and other models at 164 (38.6$\%$) stations, in average over $r=1, 2, 3$.

\begin{figure}[htb]
 	\centering
 	\begin{tabular}{l}	\includegraphics[width=13cm, height=10cm]{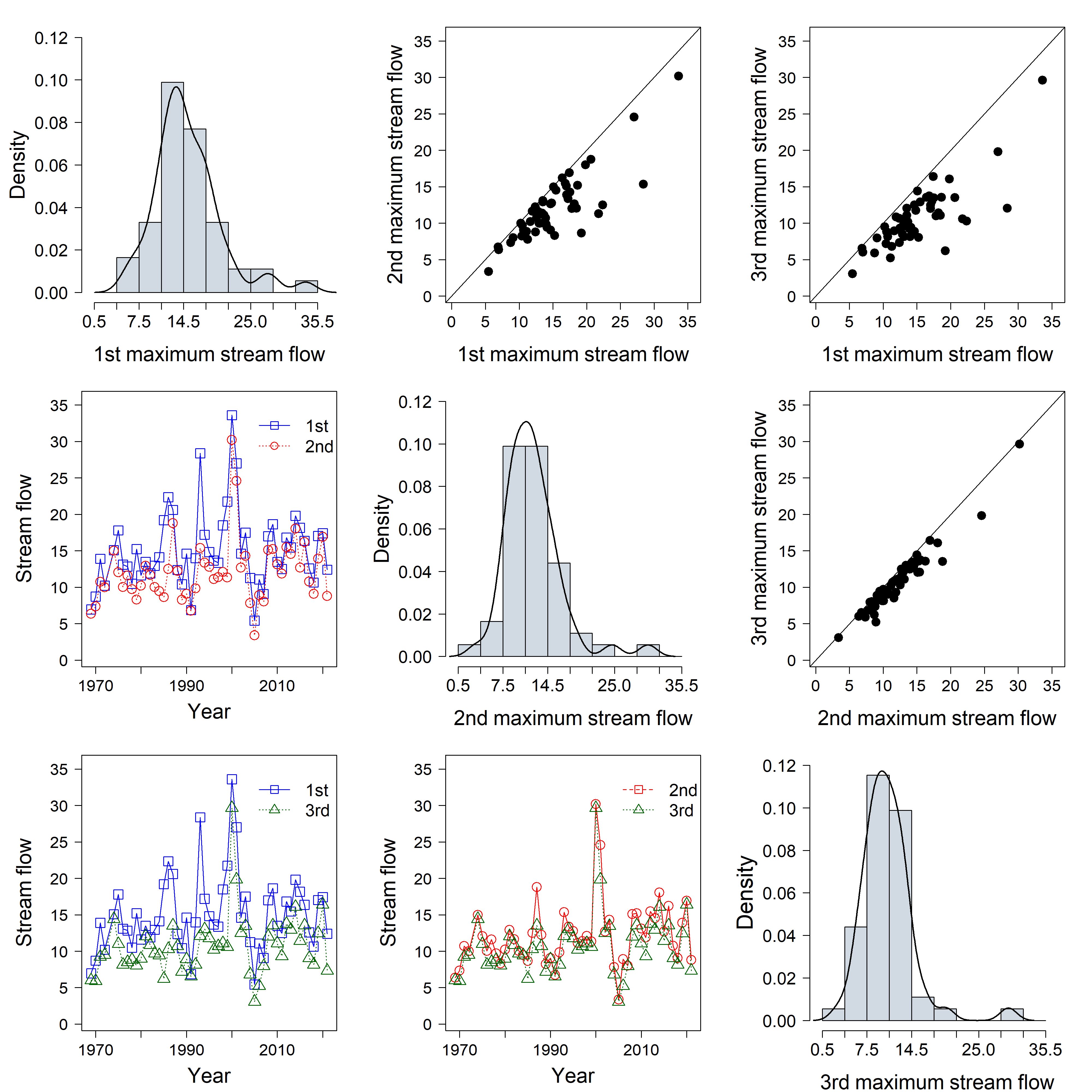} \end{tabular}	
 	\caption{Scatterplot matrix of histograms, time series plots, and scatterplots of $r$ largest order statistics for $r=1,2,3$, drawn from extreme flow data (unit: $m^3/s$ ) in Bevern Stream, UK.} \label{scatter_matrix}
 \end{figure}

  Figure~\ref{scatter_matrix} is a scatterplot matrix displaying  histograms, time series plots, and scatterplots of rLOS for $r=1,2,3$. High correlations seem to exist between consecutive-order statistics. 
 
\begin{table}[htb!]
	\centering \caption{Maximum likelihood estimates of parameters, values of negative log-likelihood (nllh) and Bayesian information criteria (BIC), and 100-year return level estimates (rl100) with standard errors in parentheses for $r=1,2,3$ obtained from the five $r$ largest models fitted to the extreme flow data (unit: $m^3/s$ ) with a sample size of $n=52$ in Bevern Stream, UK.}
	\vspace{0.4cm}
	\resizebox{0.95\textwidth}{!}{\begin{minipage}{\textwidth}
			\label{tab:BevernTable}
			\begin{tabular}{c|c|ccccccc} \hline
				model                 & $r$ & nllh  & BIC    & $\mu$       & $\sigma$       & $k$        & $h$        & rl100 (se) \\  \hline
				\multirow{3}{*}{rGLO} & 1 & 154.4 & 320.6  & 14.4 (0.63) & 2.61 (0.32) & -0.155 (0.072) &                 & 31.9 (3.6) \\
				& 2 & 254.8 & 521.4  & 14.2 (0.61) & 3.06 (0.32) & -0.174 (0.057) &                 & 35.7 (3.1) \\
				& 3 & 321.6 & 655.0  & 14.4 (0.63) & 3.27 (0.33) & -0.172 (0.053) &                 & 37.2 (2.8) \\ \hline
				\multirow{3}{*}{rK4D} & 1 & 154.3 & 324.5  & 14.8 (1.39) & 2.39 (0.67) & -0.180 (0.077) & -1.414 (1.391) & 31.8 (4.1) \\
				& 2 & 253.9 & 523.7  & 13.9 (0.61) & 3.34 (0.43) & -0.129 (0.086) & -0.519 (0.315) & 34.8 (5.2) \\
				& 3 & 320.9 & 657.7  & 14.2 (0.59) & 3.39 (0.35) & -0.149 (0.062) & -0.667 (0.257) & 36.6 (5.1) \\ \hline
				\multirow{3}{*}{rGEVD} & 1 & 155.2 & 322.3  & 12.8 (0.64) & 4.22 (0.45) & -0.042 (0.079) &                 & 30.5 (3.2) \\
				& 2 & 256.9 & 525.6  & 13.6 (0.55) & 4.24 (0.34) & -0.034 (0.061) &                 & 31.6 (3.1) \\
				& 3 & 329.4 & 670.6  & 14.2 (0.51) & 4.21 (0.30) & -0.031 (0.053) &                 & 32.2 (3.0) \\ \hline
				\multirow{3}{*}{rLD}  & 1 & 156.7 & 321.4 & 14.6 (0.64) & 2.70 (0.32) &                 &                 & 27.0 (1.4) \\
				& 2 & 259.0 & 525.9 & 14.4 (0.60) & 2.93 (0.26) &                 &                 & 27.9 (1.0) \\
				& 3 & 327.1 & 662.0 & 14.5 (0.57) & 2.96 (0.22) &                 &                 & 28.1 (0.8) \\ \hline
				\multirow{3}{*}{rGD} & 1 & 155.4 & 318.6  & 12.8 (0.61) & 4.18 (0.44) &                 &                 & 32.0 (2.3) \\
				& 2 & 257.0 & 521.9  & 13.5 (0.55) & 4.29 (0.34) &                 &                 & 33.3 (2.0) \\
				& 3 & 329.5 & 667.0    & 14.1 (0.51) & 4.29 (0.28) &                 &                 & 33.9 (1.7) \\ \hline
			\end{tabular}
	\end{minipage} }
\end{table}

\begin{figure}[htb]
	\centering
	\begin{tabular}{l}	\includegraphics[width=13cm, height=11cm]{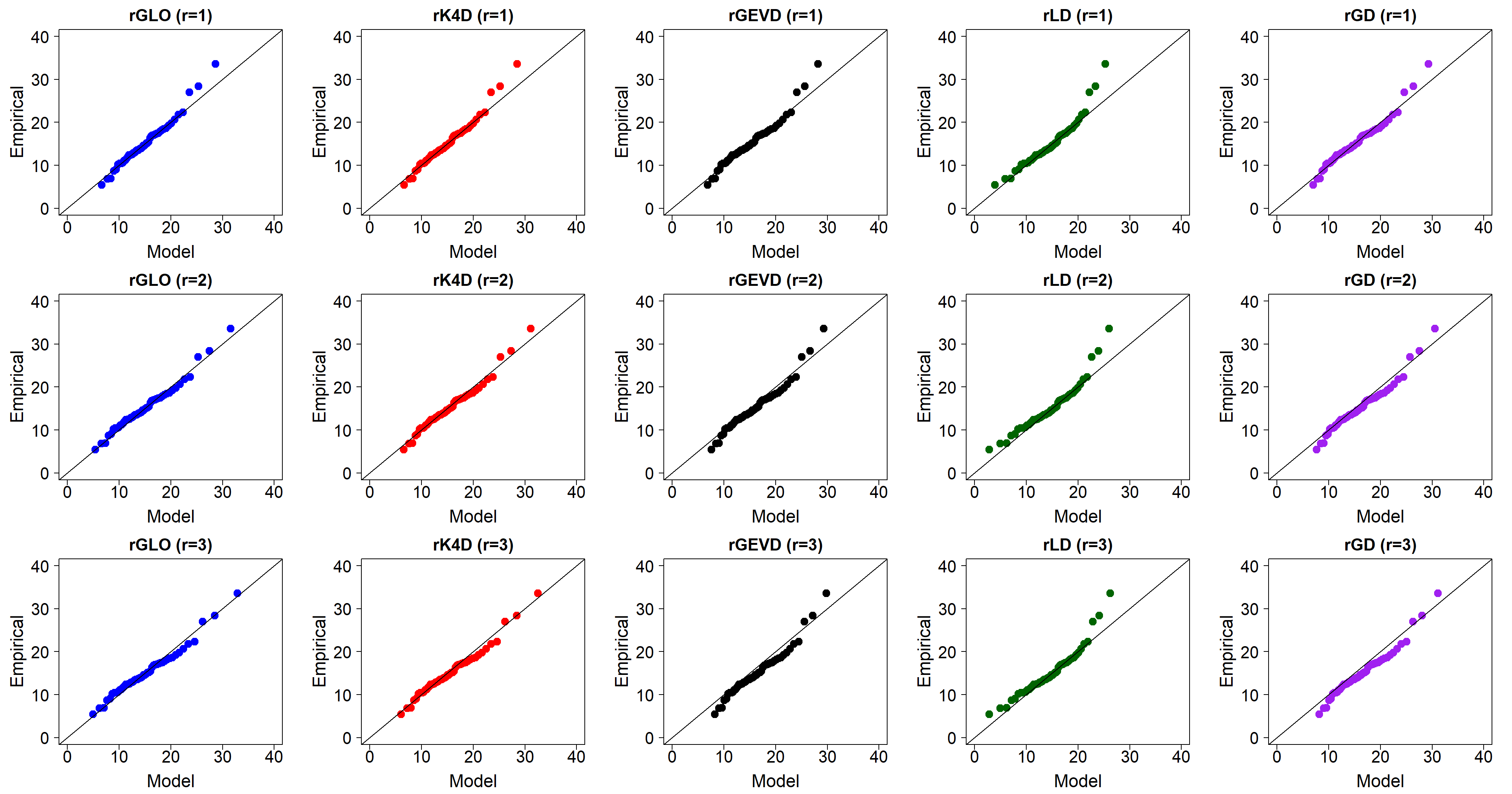} \end{tabular}	
	\caption{Quantile-per-quantile plots of the block maxima, drawn from the five $r$ largest models for $r=1,2,3$ fitted to extreme flow data with a sample size of $n=52$ in Bevern Stream, UK.} \label{qqplot}
\end{figure}

The rGLO model was fitted to the data for $r=1,2,3$. The MLE of parameters, values of negative log-likelihood (nllh), Bayesian information criteria (BIC), and 100-year return level estimates (rl100) with standard errors in parentheses for $r=1,2,3$ are given in Table~\ref{tab:BevernTable}. Similar results from the fitted rK4D, rGEVD, rLD, and rGD are also compared. Standard errors are obtained by the asymptotic formula for the MLE (i.e., inverse of Fisher information and delta method). The R package `ismev' was used for the MLE computation of the rGEVD. Figure~\ref{qqplot} displays quantile-per-quantile (Q-Q) plots of the block maxima ($s=1$) drawn from the five $r$ largest models fitted to the data. Figure~S3 in the Supplementary Material illustrates the profile log-likelihood functions and 95\% confidence intervals of the 100-year return level obtained from the five $r$ largest models.

The estimates of shape parameter $k$ from rK4D, rGLO, and rGEV are all negative, ranges from -.03 to -.18. The $k$ estimates of rGEV are close to zero and so can be reduced to rGD. Standard errors of $k$ estimate and of rl100 estimate decrease as $r$ increase from 1 to 3, for rGLO and rGEV. This SE reduction as $r$ increase from 1 to 3 suggests again to use the rGLO (or rGEV) model for $r (\ge 2)$ LOS instead of using the GLO (or GEV) distribution for just block maxima. 

The range of the $k$ estimates of rGEV is smaller (.011) than those of rGLO and rK4D (.019 and .051, respectively). In addition, the range of rl100 of rGEV is also smaller (1.7) than those of rGLO and rK4D (5.3 and 4.8, respectively). Thus the rGEV seems to provide relatively stable $k$ and rl100 estimates compared to rGLO and rK4D models. This may suggest that the rGEV model is more reliable, in the sense of stable estimation as $r$ changes from 1 to 3, for tail extrapolation. But this suggestion may be valid under the assumption that the MLE works equally well for those three models in this data. Because the MLE for K4D works much worse than the MLE for GEV (Papukdee et al.~2022), especially for negative $k$, we believe that a direct comparison between rGEV and rK4D based on only the estimated values may be an oversimplified approach. 

The SEs of rl100 of the rLD and rGD in Table~\ref{tab:BevernTable} are smaller than those of the other models. However, the Q-Q plots from the rLD and rGD in Figure~\ref{qqplot} indicate the inadequacy of the models. In the BIC and Q-Q plots, the rGD works well for $r=1$ but not for $r=3$. The rGLO has the smallest BIC among all models for all $r$, except the rGD for $r=1$. But the Q-Q plot from the rGD for $r=1$ is worse that that of the rGLO for $r=2$ or 3. The rGEVD predicts 100-year return levels lower (with comparably small SEs) than those of the rGLO and rK4D. However, the Q-Q plots of the rGEVD for $r=2$ and 3 are worse than those of the rGLO and rK4D, especially in the lower parts of the plots.

The rK4D has larger SEs in the 100-year return level estimation than the other models, which may be because of the burden in estimating the four parameters in the rK4D. Higher SE values of $h$ (and $\sigma$) estimates explain this SE inflation. Nevertheless, the Q-Q plots of the rK4D are as good as those of the rGLO. In addition, it is notable that $\hat h $ at $r=1$ is $-1.414$, which would not be obtained from the L-moment estimation because $\hat h$ is confined to be $\ge -1.0$ in the L-moment estimation of the K4D (Hosking 1994; Kjeldsen et al.~2017). 

{Table \ref{tab_bevern_ed} presents the result of the ED test for the rGLO and the rGEV for choosing an appropriate $r$. Because there are only three largest observations in Bevern streamflow data, only two testing hypothesis were performed. We found that $H_0: r=2$ and $H_0: r=3$ were accepted for both models. Thus our selection is $r=3$ based on these tests. This selection for the rGLO is consistent with the choice based on the Q-Q plot and based on SE reduction.}

\begin{table}[h!]
	\centering
	\caption{{Result of the entropy difference test for Bevern streamflow data.} }
	\vspace{0.4cm}
	\label{tab_bevern_ed}
	\begin{tabular}{cc|ccc|ccc}
		\hline
		\multicolumn{2}{c|}{hypothesis} & \multicolumn{3}{c|}{rGLO}   & \multicolumn{3}{c}{rGEV}  \\ \hline
		$H_0$             & $H_1$            & $\Bar{Y}_r$ & $T_n^{r}$ & p-value & $\Bar{Y}_r$ & $T_n^{r}$ & p-value \\ \hline
		r=2            & r=1           & -1.91  & -0.55 & 0.584    & -1.94  & 0.83 & 0.408    \\
		r=3            & r=2           & -1.28 & -0.18 & 0.858    & -1.38  & 1.35 & 0.178    \\ \hline
	\end{tabular}
\end{table}

{Table \ref{tab_cv_bic} presents values of the 5-fold CV nllh, CV AIC, and CV BIC obtained as $r$ changes from 1 to 3 when five LOS models were fitted to streamflow data in Bevern, UK. Smaller of these values are better. It turned out that the rGLO works better than the rK4D and the rGEV in three CV nllh-based criteria for all $r$. Moreover, the differences in CV nllh-based criteria between the rGLO and the rK4D are more distint than the differences in nllh-based criteria as in Table \ref{tab:BevernTable}. This is an example of justifying the advantage of the rGLO over a more general (rK4D) model. }

\begin{table}[h!]
	\centering
	\caption{{Results of the 5-fold cross-validated (CV) negative log-likelihood (nllh), CV AIC, and CV BIC obtained as $r$ changes from 1 to 3 when five $r$ largest models were fitted to Bevern streamflow data.}}
	\vspace{0.4cm}
	\label{tab_cv_bic}
	\begin{tabular}{c|ccc|ccc|ccc}
		\hline
		\multirow{2}{*}{Model} & \multicolumn{3}{c|}{$r=1$}    & \multicolumn{3}{c|}{$r=2$}      & \multicolumn{3}{c}{$r=3$}  \\ \cline{2-10}
		& cv\_nllh & cv\_aic & cv\_bic & cv\_nllh & cv\_aic & cv\_bic & cv\_nllh & cv\_aic & cv\_bic \\ \hline
		rGLO                   & 31.6     & 69.2    & 75.1    & 51.6     & 109.2   & 115.1   & 65.3     & 136.6   & 142.5   \\
		rK4D                   & 32.2     & 72.5    & 80.3    & 53.6     & 115.2   & 123.0   & 65.7     & 139.4   & 147.3   \\
		rGEV                   & 32.3     & 70.6    & 76.5    & 52.8     & 111.6   & 117.4   & 67.8     & 141.6   & 147.4   \\
		rLD                    & 32.2     & 68.4    & 72.3    & 52.6     & 109.2   & 113.0   & 66.2     & 136.5   & 140.4   \\
		rGD                    & 32.3     & 68.5    & 72.4    & 52.8     & 109.6   & 113.5   & 67.1     & 138.2   & 142.1   \\ \hline
	\end{tabular}
\end{table}

Considering all measures (the BIC, SE of parameter estimates, SE of rl100, CV likelihood criteria, and Q-Q plots) together, we conclude that the rGLO fits the extreme flow data in Bevern Stream well compared to other $r$ largest models. Our final selection of the best model for this data is the rGLO with $r=3$. 

We calculated an ensemble estimate of rl100 with a weighting scheme of the generalized L-moments distance. See  the Supplementary Material for the details of this model averaging. Actual estimate is $r_{100}^E =33.46$ obtained with $w_1 =0.63,\; w_2 =0.26,\; w_3=0.11$. To obtain the SE of this $r_{100}^E$, we applied a bootstrap method. Table \ref{tab:data_nopara_ens} are such results obtained by the nonparametric bootstrap in which the values of rl100 are the average of m=1000 bootstrap estimates. Median weights based on 1000 bootstrap samples show high weight to the first maxima, similarly to the weights used one time for real data. The SE of the ensemble is smaller than the SEs of each rGLO estimate. 
Thus we would recommend to employ a model averaging over $r$ with weights based on the generalized L-moments distance.

\begin{table}[htb!]
	\centering
	\caption{Result of nonparametric bootstrapping for $r=1,2,3$ obtained from the rGLO fitted to the extreme flow data (unit: $m^3/s$) in Bevern Stream, UK.: the 100-year return level estimates (rl100), their standard error (SE), and median weights used to calculate the ensemble estimate.}
	\label{tab:data_nopara_ens}
	\vspace{0.4cm}
	\begin{tabular}{c|ccc|c} \hline
		r     & 1    & 2    & 3    & Ensemble \\ \hline
		rl100 & 32.1 & 37.2 & 37.4 & 33.2     \\ \hline
		SE    & 4.94 & 7.71 & 7.14 & 4.75     \\ \hline
		weight & 0.765 & 0.140 & 0.052 &  \\ \hline
	\end{tabular}
\end{table}

\section{Discussion}


In the above real data study, we illustrated an example that the rGLO can serve to model the $r$ largest observations, especially when the rGEVD is not sufficient to capture the variability of observations well. Nevertheless, we do not claim that the rGLO might replace the rGEV. It is noted that one should be careful to not put too much trust in the rGLO model. But we had some evidences that the rGLO model using $r (\ge 2)$ LOS is more reliable for tail extrapolation than the GLO distribution using just block maxima ($r=1$).

We have experienced technical difficulty in finding the MLE of parameters in fitting the rGLO to Monte Carlo samples for $k > .1$ and $r > 3$. It was a divergence problem in numerically minimizing the negative log-likelihood function. The reason of failure is probably because the random sample from rGLO has too wide range including negative values, under the constraints that $ -\infty < x^{(s)} \le x^{(s-1)}\;$ and $-\infty < x \le \mu + {\frac{\sigma}{k}}$. Thus, the practical use of the rGLO based on the MLE can be limited when $k > .1$ and $r > 3$. 

When we generated a random sample ($x^{(s)}$ from the rGLO for simulation study, given $x^{(s-1)}$), some values were too low or negative especially for $k$ positive and for high $r$. Such cases do not seem consistent with real data. We thus suggest to set a lower bound constraint that $\; a \times x^{(s-1)} \le x^{(s)} \;$  with $a = .2$ for positive $k$. This constraint is more realistic than $-\infty < x^{(s)}$. For example, in Bevern Stream data, we obtained $a_2 = .45$ and $a_3 = .58$, where $\; a_s = min_i\; (x_i^{(s)}/x_i^{(s-1)})$. For Bangkok rainfall data used by Shin and Park (2023), we had $a_2=.40,\; a_3=.63,\; a_4 = .73, \; a_5=.67,\; a_6 = .56,\; a_7 =.77$, and so on. For Lowestoft sea level data used in Bader et al.~(2017), $a_2 =.70,\;a_3=.89,\;a_4=.92,\;a_5=.92,\cdots, a_{10}=.96$. For Fort Collins precipitation data in `extRemes' package (Gilleland and Katz, 2016) of R software, $a_2 =.25,\;a_3=.34,\;a_4=.52,\;a_5=.51,\; a_6=.62,\cdots, a_{10}=.77$.
Actually in Monte Carlo simulation study, we experienced much more successes in finding the MLE with the above lower bound constraint ($a = .2$ for positive $k$) than without such constraint.

Based on an investigation into the contribution of various components to the total uncertainty, Kjeldsen (2004) recommended applying the GLO with ``regional'' information to estimate high quantiles. Thus, applying the rGLO for a regional frequency analysis would be beneficial. In addition, we hope to incorporate the proposed approach with some new distributions, such as those by Stein (2021b), Yadav et al.~(2021), and Saulo et al.~(2023), to apply some ``sub-asymptotic'' extreme value models for rLOS.  

The extreme flow data in Bevern Stream seem to exhibit a slight time trend. The observed Mann-Kendall trend test statistic $\tau$ for the first maximum of Bevern Stream data is 0.148 with p-value 0.124 (no significant time trend). Non-stationary modeling in the rGLO and the rLD under changing climate are another challenge to be faced in future work.

 {In this study, we employed the entropy difference (ED) test to choose an appropriate $r$ for the rGLO model. But our approach presented here may not be sufficient. A full development and justification of new method of selecting $r$ need more work, which is a topic of future study.}


\section{Conclusion}
In this study, we introduced the GLO model for rLOS (rGLO and rLD) as special cases of the rK4D. The joint PDF and marginal and conditional distributions of rGLO were derived. The MLE, model averaging, and uncertainty quantification methods are considered to estimate the parameters and T-year return levels for the rGLO and rLD. A Monte Carlo simulation study and an application to the three largest flow data in Bevern Stream are presented to illustrate the usefulness of the rGLO. We found that the rGLO model using $r (\ge 2)$-largest order statistics performed better than the GLO distribution using just block maxima ($r=1$). {Moreover, analysis of Bevern Stream data based on some methods including the cross-validated likelihood criteria provided an example of justifying the advantage of the rGLO over a more general (rK4D) model.} In addition, it was recommended to employ a model averaging method, as an alternative to selecting the best $r$, with weights based on the generalized L-moments distance or based on the variance of return level.

The effective use of the available information is important in extremes because extreme values are scarce. Thus, using the $r$ largest method is encouraged, at least as a compromise between the block maxima and generalized Pareto models. By introducing new distributions for the $r$-largest extremes (rGLO and rLD) in this study, a pool of candidate distributions for rLOS is enriched. 



\renewcommand{\baselinestretch}{0.8}
\begin{small}    
\subsection*{Code availability}
https://github.com/yire-shin/rGLO.git

\subsection*{Supplementary Information}	
\begin{verbatim}
	https://static-content.springer.com/esm/art%3A10.1007%2Fs00477-023-02642-7/MediaObjects/ 
	477_2023_2642_MOESM1_ESM.pdf
	\end{verbatim}

\subsection*{ORCID}
Yire Shin, https://orcid.org/0000-0003-1297-5430 \\
Jeong-Soo Park, https://orcid.org/0000-0002-8460-4869

\subsection*{Acknowledgment}
	The authors would like to thank the reviewer and the Associate Editor for their valuable comments and constructive suggestions.
	This research was supported by Basic Science Research Program (No.RS-2023-00248434, 2020R1I1A3069260) and BK21 FOUR (No.5120200913674) through the National Research Foundation of Korea funded by the Ministry of Education.
\end{small}

\renewcommand{\baselinestretch}{0.8}

\end{document}